\documentclass[review]{elsarticle}
\usepackage{lineno,hyperref}
\modulolinenumbers[5]

\usepackage{amssymb}
\usepackage{mathtools}
\usepackage{amsfonts}
\usepackage{hyperref}
\usepackage[multiple]{footmisc}
\usepackage{xcolor}

\makeatletter
\def\ps@pprintTitle{%
 \let\@oddhead\@empty
 \let\@evenhead\@empty
 \def\@oddfoot{}%
 \let\@evenfoot\@oddfoot}
\makeatother

\begin{document}
\begin{frontmatter}

\title{On the weak-field limit of Pleba\'nski class electrodynamics}

\author{Gerold Oltman Schellstede\fnref{myfootnote}}
\address{ZARM, Universit{\"a}t Bremen, Am Fallturm, 28359 Bremen, Germany}
\ead{gerold.schellstede@zarm.uni-bremen.de, schellst@physik.fu-berlin.de}

\begin{abstract}
Pleba\'nski's class of nonlinear vacuum electrodynamics is considered which is for several reasons of interest at the present time. In particular the question is answered under which circumstances Maxwell's original field equations are recovered approximately and which post-Maxwellian effects could arrise. To this end a weak field approximation method is developed allowing to calculate post-Maxwellian corrections up to Nth order. In some respect this is analogue of determining "post-newtonian" corrections from relativistic mechanics by a low velocity approximation. As a result we got a series of linear field equations which can be solved order by order. In this context the solutions of the lower orders occur as source terms inside the higher order field equations and represent a "post-Maxwellian" self-interaction of the electromagnetic field which increases order by order. One has to decide between problems with and without external source-terms, because without also high frequency solutions can be approximately described by Maxwell's original equations. The higher order approximations which describe "post-Maxwellian" effects can give rise for experimental tests of Pleba\'nksi's class. Finally two boundary value problems are discussed to have examples at hand.
\end{abstract}

\begin{keyword}
Nonlinear vacuum electrodynamics \sep Pleba\'nksi class \sep Born-Infeld theory \sep Heisenberg-Euler theory \sep Weak-field limit \sep post-Maxwell approximation \sep Boundary value problems
\end{keyword}

\end{frontmatter}

\linenumbers

\section{Introduction}
Attempts to modify Maxwell's fundamental theory of electromagnetism have a long history~\cite{Sommerfeld} since in 1912 Gustav Mie developed such a modification to overcome some problems as infinite self-energies and self-forces. While Mie's theory~\cite{Mie1,Mie2} suffers from serious problems -- e.g. the gauge invariance is broken -- it stimulated the research on this field enduringly. The first reasonable theory which could prevent infinite self-energies was Born's one~\cite{MB} presented in 1933 which experienced further development together with Infeld in 1934 and led to Born-Infeld's theory~\cite{BornInfeld1934}. On the other hand in 1936 Heisenberg and Euler developed a theory with the different aim to incorporate some effects of quantum electrodynamics into an effective classical electromagnetic theory~\cite{HeisenbergEuler1936,GD}. 

Both Heisenberg-Euler's theory as well as Born's respectively Born-Infeld's theory can be developed from a U(1)-gauge and Lorentz invariant Lagrangian of the form $L=L(F,G)$ (compare (\ref{Feldinvarianten})). This class is known as Pleba\'nski class and was comprehensively discussed by Pleba{\'n}ski~\cite{Plebanski1970} and Boillat~\cite{Boillat}. Recently this class of theories gained attention from string theory because following Tseytlin~\cite{Tseytlin1999} Born-Infeld's theory can be viewed as an effective electrodynamics stemming from some versions of string theory. 

In the present article we answer among which conditions Maxwell's original theory is approximately recovered and which post-Maxwellian corrections arise.

To do so an approximation procedure for the whole class is developed. This means that a weak-field approximation up to Nth order is performed and compared with Maxwell's theory. This enables one to find post-Maxwellian terms which can be viewed as a starting point to experimental verifications for Pleba{\'n}ski's class.\footnote{Here it is worth to mention that for Born-Infeld's theory there is also an indirect approximation method for two-dimensional problems recently discussed in~\cite{Ferraro}.}

The article is organized as follows: In section \ref{section2} we perform the weak field series expansions for the field strength as well as for the field invariant $F$ and the pseudo-invariant $G$. This enables one then to do this also for the derivatives of the Lagrangian with respect to the invariants $L_F:=\partial L/\partial F$ and $L_G:=\partial L/\partial G$. In section \ref{section3} the discussion of the field equations is performed and the question of gauge invariance is tackled. Furthermore we discuss how to treat boundary values in the context of our approximation. In section \ref{section4} some boundary value problems are solved with the help of our approximate field equations, while the conclusions are drawn in section \ref{section5}. 

The following conventions are used in this paper: Latin indices run from $1$ to $4$ while Greek one run from $1$ to $3$. For the sake of simplicity Gaussian CGS-units are used. A Minkowski space-time is presumed with $\eta^{ab}=\text{diag}[1,1,1,-1]$ and $\varepsilon^{1234}=-1$. 

For further use the following relations should be noted too. The hodge dual of the electromagnetic field tensor is given by 
\begin{eqnarray}
\tilde F^{mn}=\frac{1}{2}\,\varepsilon^{mnab}F_{ab}&\Rightarrow&\tilde{\tilde F}{}^{mn}=\frac{1}{2}\,\varepsilon^{mnab}\tilde F_{ab}=-F^{mn}\,,
\end{eqnarray}
The field invariant $F$ and the pseudo-invariant $G$ are defined by
\begin{eqnarray}\label{Feldinvarianten}
F=\frac{1}{2}\,F_{mn}F^{mn}=-\frac{1}{2}\,\tilde F{}_{mn}\tilde F{}^{mn}\,;\quad G=-\frac{1}{4}\,F_{mn}\tilde F^{mn}\,,
\end{eqnarray}
with
\begin{equation}
\frac{\partial F}{\partial F_{ab}}=2F^{ab}\,;\quad\frac{\partial F}{\partial \tilde F_{ab}}=-2\tilde F^{ab}\,;\quad\frac{\partial G}{\partial F_{ab}}=-\tilde F^{ab}\,.
\end{equation}
Sometimes we will make use of the relation $\varepsilon_{abcd}\varepsilon^{cdmn}=-2\left(\delta^m_c\delta^n_d-\delta^m_d\delta^n_c\right)$.

\section{Preliminaries for the approxmation}\label{section2}
\subsection{Hamilton's principle and field equations }
The field equations follow from a variational principle where the vector potential $A_m$ is used to secure the fulfillment of the homogeneous Maxwell equation
\begin{equation}\label{MaxwellHom}
\partial_{<a}F_{bc>}=0\quad\Leftrightarrow\quad\partial_b\tilde{F}^{ab}=0\quad\Leftrightarrow\quad F_{mn}=\partial_mA_n-\partial_nA_m\,
\end{equation}
where $A_m$ is only determined up to an arbitrary $U(1)$ gauge transformation
\begin{equation}
A_m\,\rightarrow\,A_m+\partial_m\chi\,.
\end{equation}

The variation of the action integral defined by
\begin{equation}
S[A_m]=\frac{1}{4\pi\,c}\int \left(  L(F_{mn})+\frac{4\pi}{c}\,j^mA_m\right)\mathrm dV_4\,
\end{equation}
leads to the inhomogeneous field equation which generally differs from Maxwell's original one
\begin{equation}
\partial_b\left(-\frac{\partial  L}{\partial F_{ab}}\right)=\frac{4\pi}{c}\,j^a\,.
\end{equation}

In the language of electrodynamics in media, the vacuum offers a nonlinear constitutive law being local in time and space while the Lagrangian density is an arbitrary function of the field strength. Therefore, usually one defines the excitation as follows (compare e. g.~\cite{Zwiebach})
\begin{equation}
H^{ab}:=-\frac{\partial  L}{\partial F_{ab}}\quad\Leftrightarrow\quad \tilde H{}^{ab}:=\frac{\partial  L}{\partial \tilde F_{ab}}\,.
\end{equation}
Thus Maxwell's equations for field strength and excitation are also valid in nonlinear electrodynamics, but the linear constitutive law for the vacuum $H^{mn}=F^{mn}$ known from Maxwell's theory is substituted by the nonlinear one given above.

Until now the Lagrangian was an arbitrary function of the field strength. The Pleba\'nski class considered here restricts the Lagrangian to a Lorentz-invariant but not parity invariant function of the field strength by setting it as an arbitrary function of the field invariant $F$ and the pseudo-invariant $G$~\cite{Plebanski1970}. The inhomogeneous field equation of the Pleba\'nski class type ($L=L(F,G)$) reads then 
\begin{equation}\label{MaxwellInhom}
\partial_bH^{ab}=\frac{4\pi}{c}\,j^a\quad\text{with}\quad H^{ab}=-2\,L_F\,F^{ab}+L_G\,\tilde F^{ab}\,.
\end{equation}

\subsection{Series expansions}
The theories pooled as Pleba\'nski's class introduce a new constant of nature $A>0$ with the unit of a field strength so that the Lagrangian can be written as a function of two dimensionless arguments
\begin{equation}
 L=L\left(\frac{F}{A^2},\frac{G}{A^2}\right)\,.
\end{equation} 
Born-Infeld's theory for example requires always $1+F/A^2-G^2/A^4>0\,\Rightarrow\, F>-A^2$ ($A\equiv b_0$) what results in electrostatic situations ($\mathbf B=0$) in an upper bound for the electric field strength~\cite{BornInfeld1934}. One can compare the situation to relativistic mechanics where $c$ is the upper bound for the particle velocity $v$. 

In analogy to the approach in~\cite{Sch} we make a series expansion with respect to $\epsilon\sim F^{mn}/A$ to tackle the question under which conditions a theory of Pleba\'naki's class is compatible with Maxwell's theory as a weak field limit and to derive post-Maxwellian terms for the field equations. To do so some series expansions are needed which will be done next. 

To this end the field strength and its hodge dual is represented as a series in the parameter of smallness $\epsilon$ 
\begin{equation}\label{analog}
F^{mn}=\sum\limits_{N=0}^\infty \epsilon^N \underset{N}f^{mn}\quad\Leftrightarrow\quad\tilde{F}_{mn}=\sum\limits_{N=0}^\infty \epsilon^N \underset{N}{\tilde f}{}_{mn}\,.
\end{equation}
In doing so the validity of the following bookkeeping system is assumed
\begin{equation}\label{analog2}
\underset{N}f^{mn}\sim \underset{N}{\tilde f}{}_{mn}\sim\underset{N+1}f^{mn}\sim\underset{N+1}{\tilde f}{}_{mn}\quad\text{and}\quad\epsilon\sim\underset{N}{f}{}_{mn}/A\quad\,.
\end{equation}

In analogy to (\ref{analog}) a series for the excitation $H^{mn}$ is needed. Therefore a series for the derivatives $L_F$ and $L_G$ of the Lagrangian in $F$ and $G$ as well as series for the field invariants $F$, $G$ and their mixed powers $F^MG^N$ have to be calculated. 

We start with the series for $F$ and $G$:
\begin{equation}
\begin{split}
&F=\sum\limits_{N=0}^{\infty}\underset{N}F\,\epsilon^N\quad\text{with}\quad \underset{N}F=\frac{1}{2}\sum\limits_{K=0}^N \underset{K}{f}{}_{mn}\underset{N-K}{{f}{}^{mn}}\,,\\
&G=\sum\limits_{N=0}^{\infty}\underset{N}G\,\epsilon^N\quad\text{with}\quad \underset{N}G=-\frac{1}{4}\sum\limits_{K=0}^N \underset{K}{f}{}_{mn}\underset{N-K}{\tilde f}{}^{mn}\,.
\end{split}
\end{equation}
As shown in appendix \ref{PotenzenVonPotenzreihen} one gets for the powers of them 
\begin{equation}
\begin{split}
&F^N=\sum\limits_{J=0}^\infty\underset{J}{\overset{N}F}\,\epsilon^J\\
&\text{with}\quad \underset{J}{\overset{N}F}=\sum\limits_{s}^S\frac{N\cdot(N-1)\cdot\ldots\cdot(N-M+1)}{k_{1_s}!\cdot\ldots\cdot k_{J_s}!}\,\underset{0}F^{N-M}\underset{1}F^{k_{1_s}}\cdot\ldots\cdot \underset{J}F^{k_{J_s}}\,,\\
&G^N=\sum\limits_{J=0}^\infty\underset{J}{\overset{N}G}\,\epsilon^J\\
&\text{with}\quad \underset{J}{\overset{N}G}=\sum\limits_{s}^S\frac{N\cdot(N-1)\cdot\ldots\cdot(N-M+1)}{k_{1_s}!\cdot\ldots\cdot k_{J_s}!}\,\underset{0}G^{N-M}\underset{1}G^{k_{1_s}}\cdot\ldots\cdot \underset{J}G^{k_{J_s}}\,,
\end{split}
\end{equation}
where one sums up over all solutions of 
\begin{equation}
k_{1_s}+2k_{2_s}+\ldots+Jk_{J_s}=J\quad\text{with}\quad M:=k_{1_s}+k_{2_s}+\ldots+k_{J_s}\leqq N\,
\end{equation}
numbered by $s$ whose total number is $S$.

This leads to the following series representation of the mixed powers 
\begin{equation}\label{ProduktFG}
\left(\frac{F}{A^2}\right)^{N-K}\left(\frac{G}{A^2}\right)^{K}=\frac{1}{A^{2N}}\sum\limits_{J=0}^\infty\left(\sum\limits_{T=0}^J \underset{T}{\overset{K}G}\,\underset{J-T}{\overset{N-K}F}\right)\epsilon^J\,.
\end{equation}
As a consequence of our bookkeeping system (\ref{analog2}) it follows
\begin{equation}
\underset{J}F/A^2\sim\underset{N}{f}{}_{mn}\underset{P}{f}{}^{mn}/A^2\sim\epsilon^2\quad\Rightarrow\quad\underset{J}{\overset{N}F}/A^{2N}\sim\epsilon^{2N}\,.
\end{equation}

With this we calculate a series for $L_F$ and $L_G$ whereat we treat them together as $L_X$ with $X=F,G$. First we develop $L_X$ as MacLaurin series in $F$ and $G$
\begin{equation}
\begin{split}
L_X&=\sum\limits_{N=0}^\infty\frac{1}{N!}\left(\frac{F}{A^2}\frac{\partial}{\partial (F/A^2)}+\frac{G}{A^2}\frac{\partial}{\partial (G/A^2)}\right)^NL_X(0,0)\\
&=\sum\limits_{N=0}^\infty\frac{1}{N!}\left(\sum\limits_{K=0}^N\left(\begin{array}{c}
N\\K\end{array}\right)\left(F/A^2\right)^{N-K}\left(G/A^2\right)^{K}\times\right.\\
&\qquad\qquad\qquad\qquad\left.\times\left.\frac{\partial^{N-K}}{\partial (F/A^2)^{N-K}}\frac{\partial^{K}}{\partial (G/A^2)^{K}}L_X\right|_{F=0,G=0}\right)\,,
\end{split}
\end{equation}
second we insert the product (\ref{ProduktFG}) and get
\begin{equation}
\begin{split}
L_X=&\sum\limits_{J=0}^\infty\sum\limits_{N=0}^\infty\left(\frac{1}{A^{2N}}\sum\limits_{K=0}^N\sum\limits_{T=0}^J\frac{1}{N!}\left(\begin{array}{c}
N\\K\end{array}\right)\underset{T}{\overset{K}G}\,\underset{J-T}{\overset{N-K}F}\times\right.\\
&\qquad\qquad\qquad\qquad\left.\times\left.\frac{\partial^{N-K}}{\partial (F/A^2)^{N-K}}\frac{\partial^{K}}{\partial (G/A^2)^{K}}L_X\right|_{F=0,G=0}\right)\epsilon^J\,.
\end{split}
\end{equation}
Here one has to consider that the order in $\epsilon$ of each summand depends not only on $J$ but also on $N$. In fact $J$ indicates the lowest possible order which is raised due to $N>0$. However, for our approach it is indispensable to know which summands of this series are of the same order $Q$. One sees that all summands which fulfill $2N+J=:Q$ are of the fixed but arbitrary order $Q$. To underline this we write\footnote{The symbol $\sum_{2N+J=Q}$ means to sum up over all solutions $N,J$ of $2N+J=Q$ for a given $Q$.}
\begin{equation}
\begin{split}
L_X=&\sum\limits_{Q=0}^\infty\left(\sum\limits_{2N+J=Q}\sum\limits_{K=0}^N\sum\limits_{T=0}^J\frac{1}{\epsilon^{2N}}\frac{1}{A^{2N}}\frac{1}{N!}\left(\begin{array}{c}
N\\K
\end{array}\right)\underset{T}{\overset{K}G}\,\underset{J-T}{\overset{N-K}F}\times\right.\\
&\qquad\qquad\qquad\qquad\left.\times\left.\frac{\partial^{N-K}}{\partial (F/A^2)^{N-K}}\frac{\partial^{K}}{\partial (G/B^2)^{K}}L_X\right|_{F=0,G=0}\right)\epsilon^Q\\
=&\sum\limits_{Q=0}^\infty\underset{Q}L{}_X\,\epsilon^Q\,.
\end{split}
\end{equation}
It is clear that the derivatives of $L_X$ at $F=G=0$ are plainly numbers. If the Lagrangian of some specific theory under consideration is given as a power series in $F$ and $G$ they are easily related to the coefficients of that series. 

This enables one to write down the product series of the field strength $F^{mn}$ or its dual $\tilde{F}^{mn}$ with $L_F$ or $L_G$ (compare appendix \ref{ProduktPotenzreihen}). To ease things up the combined notation $Y^{mn}=F^{mn},\tilde{F}^{mn}$ is used by which one gets
\begin{equation}\label{LXY}
\begin{split}
&L_{X}Y^{mn}=\sum\limits_{P=0}^\infty\left(\sum\limits_{Q=0}^P\underset{Q}L{}_X\underset{P-Q}Y^{mn}\right)\epsilon^P\\
\Rightarrow\,\,&L_{X}Y^{mn}=\sum\limits_{P=0}^\infty\left\{\sum\limits_{Q=0}^P\left(\sum\limits_{2N+J=Q}\sum\limits_{K=0}^N\sum\limits_{T=0}^J\frac{1}{\epsilon^{2N}}\frac{1}{A^{2N}}\frac{1}{N!}\left(\begin{array}{c}
N\\K
\end{array}\right)\underset{T}{\overset{K}G}\underset{J-T}{\overset{N-K}F}\right.\right.\times\\
&\qquad\qquad\qquad\times\left.\left.\left.\frac{\partial^{N-K}}{\partial (F/A^2)^{N-K}}\frac{\partial^{K}}{\partial (G/A^2)^{K}}L_X\right|_{F=0,G=0}\right)\underset{P-Q}Y^{mn}\right\}\epsilon^P\,.
\end{split}
\end{equation}

This enables us to write down the excitation as a series in $\epsilon$.
\begin{equation}\label{ApproxErregung}
\begin{split}
H^{mn}=\,&-2L_FF^{mn}+L_G\tilde{F}^{mn}\\
=\,&\sum\limits_{P=0}^\infty\left(-2\sum\limits_{Q=0}^P\underset{Q}L{}_F\underset{P-Q}f^{mn}+\sum\limits_{Q=0}^P\underset{Q}L{}_G\underset{P-Q}{\tilde{f}}{}^{mn}\right)\epsilon^P\\
=\,&\sum\limits_{P=0}^\infty\underset{P}h{}^{mn}\,\epsilon^P\,.
\end{split}
\end{equation}

Maxwell's inhomogeneous field equation (\ref{MaxwellInhom}) needs also a series for the current $j^m$ on its right side with the same parameter of smallness $\epsilon$
\begin{equation}
j^m=\sum\limits_{P=0}^\infty \underset{P}j^m\epsilon^P\quad\text{with}\quad \underset{P}j^m\sim\underset{P+1}j^m\,.
\end{equation}

At last one has to estimate the impact of the derivative onto the excitation or indirectly onto the field strength. Because our consideration should not be restricted to low frequencies a second parameter $\tau$ has to be introduced
\begin{equation}\label{tau}
\partial_a\underset{N}{f}{}_{mn}\sim\tau\underset{N}{f}{}_{mn}\,.
\end{equation}

\section{Approximation of the field equations}\label{section3}
\subsection{The homogeneous Maxwell equation}\label{TheHomogeneousMaxwellEquation}
Inserting the series of the dual field strength into the homogeneous Maxwell equations (\ref{MaxwellHom}) one obtains
\begin{equation}\label{FG_HOM}
\begin{split}
\partial_n\tilde{F}^{mn}=0\quad\Rightarrow\quad\sum\limits_{N=0}^\infty\epsilon^N\partial_n\underset{N}{\tilde{f}}{}^{mn}=0\quad\Rightarrow\quad\partial_n\underset{N}{\tilde{f}}{}^{mn}=0\,.
\end{split}
\end{equation}
Each coefficient of the series for the dual field strength fulfills the homogeneous Maxwell equation separately where the magnitude of the parameter $\tau$ needs not to be specified, so its relation to $\epsilon$ remains open. 

Obviously one can fulfill the homogeneous field equation order by order with aid of a potential $\underset{N}A{}_n$ where comparison with (\ref{MaxwellHom}) shows the relation between  $A_n$ and $\underset{N}A{}_n$
\begin{equation}\label{Potential_N}
\underset{N}f{}_{mn}=\partial_m\underset{N}A{}_n-\partial_n\underset{N}A{}_m\quad\Rightarrow\quad A_n=\sum\limits_{N=0}^\infty\epsilon^N\underset{N}A{}_n\,.
\end{equation} 

In principle this offers the possibility to assume different gauges conditions for every order of the potential. This is similar to the situation known from shockwaves in Pleba{\'n}ski's class~\cite{Minz}.

\subsection{The inhomogeneous Maxwell equation}\label{TheInhomogeneousMaxwellEquation}
The difference between the theories of the Pleba\'nski class and the special subcase of Maxwell's theory is related to different constitutive laws (\ref{ApproxErregung}) for the vacuum whereas the field equations for the field strength and the excitation remain unchanged. 

The leading coefficients of the excitation series (\ref{ApproxErregung}) read
\begin{eqnarray}
\underset{0}h^{mn}&=&-2\left.L_F\right|_{F=0,G=0}\underset{0}{f}{}^{mn}+\left.L_G\right|_{F=0,G=0}\underset{0}{\tilde f}{}^{mn}\\
\underset{1}h^{mn}&=&-2\left.L_F\right|_{F=0,G=0}\underset{1}{f}{}^{mn}+\left.L_G\right|_{F=0,G=0}\underset{1}{\tilde f}{}^{mn}\\
\underset{2}h^{mn}&=&-2\left.L_F\right|_{F=0,G=0}\underset{2}{f}{}^{mn}+\left.L_G\right|_{F=0,G=0}\underset{2}{\tilde f}{}^{mn}\\
\nonumber&&-\frac{2}{\epsilon^2}\left(\frac{\underset{0}F}{A^2}\left.\frac{\partial L_F}{\partial (F/A^2)}\right|_{F=0,G=0}+\frac{\underset{0}G}{A^2}\left.\frac{\partial L_F}{\partial (G/A^2)}\right|_{F=0,G=0}\right)\underset{0}{f}{}^{mn}\\
\nonumber&&+\frac{1}{\epsilon^2}\left(\frac{\underset{0}F}{A^2}\left.\frac{\partial L_G}{\partial (F/A^2)}\right|_{F=0,G=0}+\frac{\underset{0}G}{A^2}\left.\frac{\partial L_G}{\partial (G/A^2)}\right|_{F=0,G=0}\right)\underset{0}{\tilde f}{}^{mn}
\end{eqnarray}
\begin{eqnarray}
\underset{3}h^{mn}&=&-2\left.L_F\right|_{F=0,G=0}\underset{3}{f}{}^{mn}+\left.L_G\right|_{F=0,G=0}\underset{3}{\tilde f}{}^{mn}\\
\nonumber&&-\frac{2}{\epsilon^2}\left(\frac{\underset{0}F}{A^2}\left.\frac{\partial L_F}{\partial (F/A^2)}\right|_{F=0,G=0}+\frac{\underset{0}G}{A^2}\left.\frac{\partial L_F}{\partial (G/A^2)}\right|_{F=0,G=0}\right)\underset{1}{f}{}^{mn}\\
\nonumber&&-\frac{2}{\epsilon^2}\left(\frac{\underset{1}F}{A^2}\left.\frac{\partial L_F}{\partial (F/A^2)}\right|_{F=0,G=0}+\frac{\underset{1}G}{A^2}\left.\frac{\partial L_F}{\partial (G/A^2)}\right|_{F=0,G=0}\right)\underset{0}{f}{}^{mn}\\
\nonumber&&+\frac{1}{\epsilon^2}\left(\frac{\underset{0}F}{A^2}\left.\frac{\partial L_G}{\partial (F/A^2)}\right|_{F=0,G=0}+\frac{\underset{0}G}{A^2}\left.\frac{\partial L_G}{\partial (G/A^2)}\right|_{F=0,G=0}\right)\underset{1}{\tilde f}{}^{mn}\\
\nonumber&&+\frac{1}{\epsilon^2}\left(\frac{\underset{1}F}{A^2}\left.\frac{\partial L_G}{\partial (F/A^2)}\right|_{F=0,G=0}+\frac{\underset{1}G}{A^2}\left.\frac{\partial L_G}{\partial (G/A^2)}\right|_{F=0,G=0}\right)\underset{0}{\tilde f}{}^{mn}\,
\end{eqnarray}
and so on (the fourth and fifth order can be found in appendix \ref{FuenfteVierteOrd}). One sees immediately that in every even order new mathematical structures come up while the following odd order only enriches them.  

Consequently the only possibility to recover Maxwell's theory in a certain order of approximation is given if the zeroth and the first orders are in agreement with Maxwell. Here one has to distinguish the vacuum case with $j^m=0$ and the general case with an arbitrary $j^m$.

In the vacuum case for arbitrary  $\left.L_F\right|_{F=0,G=0}\neq0$ and $\left.L_G\right|_{F=0,G=0}$ one gets Maxwell's original equations in the zeroth and first orders
\begin{eqnarray}
\partial_n\underset{0}{\tilde f}{}^{mn}=0&\quad\Rightarrow\quad&\partial_n\underset{0}{f}{}^{mn}=0\,,\\
\partial_n\underset{1}{\tilde f}{}^{mn}=0&\quad\Rightarrow\quad&\partial_n\underset{1}{f}{}^{mn}=0\,.
\end{eqnarray}
Interestingly the $\left.L_G\right|_{F=0,G=0}$-terms have no influence on the field equations up to this order which means that Maxwell's field equations can be recovered even if Maxwell's constitutive law $H^{mn}=F^{mn}$ is not recovered, because instead of $\underset{0/1}h^{mn}=\underset{0/1}{f}{}^{mn}$ one has generally
\begin{equation}\label{Erregung_Test}
\underset{0/1}h^{mn}=-2\left.L_F\right|_{F=0,G=0}\underset{0/1}{f}{}^{mn}+\left.L_G\right|_{F=0,G=0}\underset{0/1}{\tilde f}{}^{mn}\,.
\end{equation}
If $H^{mn}$ in nonlinear electrodynamics would be measurable like in standard electrodynamics in media this would provide a possibility for tests of Pleba\'nski's class.

Another remarkable point is that in vacuum ($j_m=0$) no prediction for the magnitude of $\tau$ is necessary. This means that all Pleba\'nski class electrodynamics of $\left.L_F\right|_{F=0,G=0}\neq0$ type recover Maxwell's standard field equations as limiting case for weak fields regardless if these weak fields are of low or high frequency.

For general $j^m$ both is no longer true. One has to choose $\left.L_F\right|_{F=0,G=0}$ and $\tau$ in a way that the zeroth and first orders recover Maxwell's theory with current
\begin{eqnarray}
\partial_n\underset{0}{\tilde f}{}^{mn}=0&\quad\Rightarrow\quad&\partial_n\underset{0}{f}{}^{mn}=\frac{4\pi}{c}\,\underset{0}j^m\,,\\
\partial_n\underset{1}{\tilde f}{}^{mn}=0&\quad\Rightarrow\quad&\partial_n\underset{1}{f}{}^{mn}=\frac{4\pi}{c}\,\underset{1}j^m\,.
\end{eqnarray}
This is only achievable for the choice 
\begin{equation}\label{tau2}
\left.L_F\right|_{F=0,G=0}=-\frac{1}{2}\,;\quad\tau\underset{N}{f}{}_{mn}\sim\frac{4\pi}{c}\,\underset{N}j^m\,.
\end{equation}
Physically this implies an additional restriction to low frequencies which means that for a general $j^m\neq0$ Maxwell's theory can only be recovered for cases where ($\ref{tau2}_{\mathrm{II}}$) is fulfilled. As stated before only the non-vacuum case with $j^m\neq0$ is restricted in this way while the vacuum case recovers Maxwell's theory for arbitrary frequencies.

This assumption for $\tau$ secures also that in case of a non-vanishing current the same order of the left side of the inhomogeneous field equations couples to the same order of the right side which means
\begin{equation}\label{FG_Inhom}
\partial_nH^{mn}=\frac{4\pi}{c}\,j^m\quad\Rightarrow\quad \partial_n\underset{N}h^{mn}=\frac{4\pi}{c}\,\underset{N}j^m\,
\end{equation}

Here the equations ($(\ref{FG_Inhom}_{\text{II}})$ are linear in $\underset{N}f^{mn}$ and it is worth to discuss the second and third orders equations for $N=2$ and $N=3$
\begin{equation}
\begin{split}
&\partial_n\left(\epsilon^2\underset{2}{f}{}^{mn}\right)=-\frac{1}{\left.L_F\right|_{F=0,G=0}}\times\\
&\,\,\,\times\left\{\partial_n\left(\frac{\underset{0}F}{A^2}\left.\frac{\partial L_F}{\partial (F/A^2)}\right|_{F=0,G=0}\underset{0}{f}{}^{mn}+\frac{\underset{0}G}{A^2}\left.\frac{\partial L_F}{\partial (G/A^2)}\right|_{F=0,G=0}\underset{0}{f}{}^{mn}\right)\right.\\
&\,\,\qquad-\left.\frac{1}{2}\underset{0}{\tilde f}{}^{mn}\partial_n\left(\frac{\underset{0}F}{A^2}\left.\frac{\partial L_G}{\partial (F/A^2)}\right|_{F=0,G=0}+\frac{\underset{0}G}{A^2}\left.\frac{\partial L_G}{\partial (G/A^2)}\right|_{F=0,G=0}\right)\right\}\\
&\,\,\,-\frac{1}{\left.L_F\right|_{F=0,G=0}}\frac{2\pi}{c}\epsilon^2\underset{2}j^m\,
\end{split}
\end{equation}
and 
\begin{equation}
\begin{split}
&\partial_n\left(\epsilon^3\underset{3}{f}{}^{mn}\right)=-\frac{\epsilon}{\left.L_F\right|_{F=0,G=0}}\times\\
&\,\,\,\times\left\{\partial_n\left(\frac{\underset{0}F}{A^2}\left.\frac{\partial L_F}{\partial (F/A^2)}\right|_{F=0,G=0}\underset{1}{f}{}^{mn}+\frac{\underset{0}G}{A^2}\left.\frac{\partial L_F}{\partial (G/A^2)}\right|_{F=0,G=0}\underset{1}{f}{}^{mn}\right)\right.\\
&\,\,\qquad+\partial_n\left(\frac{\underset{1}F}{A^2}\left.\frac{\partial L_F}{\partial (F/A^2)}\right|_{F=0,G=0}\underset{1}{f}{}^{mn}+\frac{\underset{1}G}{A^2}\left.\frac{\partial L_F}{\partial (G/A^2)}\right|_{F=0,G=0}\underset{0}{f}{}^{mn}\right)\\
&\,\,\qquad-\frac{1}{2}\underset{1}{\tilde f}{}^{mn}\partial_n\left(\frac{\underset{0}F}{A^2}\left.\frac{\partial L_G}{\partial (F/A^2)}\right|_{F=0,G=0}+\frac{\underset{0}G}{A^2}\left.\frac{\partial L_G}{\partial (G/A^2)}\right|_{F=0,G=0}\right)\\
&\,\,\qquad-\left.\frac{1}{2}\underset{0}{\tilde f}{}^{mn}\partial_n\left(\frac{\underset{1}F}{A^2}\left.\frac{\partial L_G}{\partial (F/A^2)}\right|_{F=0,G=0}+\frac{\underset{1}G}{A^2}\left.\frac{\partial L_G}{\partial (G/A^2)}\right|_{F=0,G=0}\right)\right\}\\
&\,\,\,-\frac{1}{\left.L_F\right|_{F=0,G=0}}\frac{2\pi}{c}\epsilon^3\underset{3}j^m\,.
\end{split}
\end{equation}
Here the zeroth order solution acts as a nonlinear source term from the second order on whereas the first order solution occurs as source term from the third order on. One sees clearly that the third order brings no new mathematical structures into play but enriches the existing ones. 

Looking at the excitations of fourth and fifth orders (compare appendix \ref{FuenfteVierteOrd}) it is seen that again the fourth order brings new terms into play whereas the fifth order enriches the structure. For the field equations of fourth and fifth orders this means that the solutions of the second and third orders enriches the source terms introduced in the second and third orders, where the solutions of the zeroth and first orders build new source terms with a higher degree of nonlinearity. The result is that from the second order on a self-interaction of the electromagnetic field comes into play and this effect increases with higher orders by adding higher powers of the field invariants as sources. 

At this point we would make a remark about boundary value problems in view of the presented approximation. In principle boundary values can be also given as a series in $\epsilon$. Then every order has to fulfill the boundary value of the correspondent order. However, a second approach is possible where the zeroth order fulfills the boundary values and all higher orders fulfill natural boundary values which means they vanish at the boundary. In both versions the whole series for the field strength fulfills the same values on the boundary. We will see in the next section that this is not of further interest for practical applications. 

\subsection{Combined form of approximate field equations}\label{CombinedFormOfApproximateFieldEquations}
For practical applications the approximate field equations are not very convenient. Therefore we derive an alternative variant to calculate the field up to a desired order. To do so we introduce the following new quantities
\begin{equation}
\begin{split}
&\underset{*N}{f}{}^{mn}:=\sum\limits_{J=0}^N \epsilon^J\underset{J}{f}{}^{mn}\,;\quad\underset{*N}{\tilde f}{}^{mn}:=\sum\limits_{J=0}^N \epsilon^J\underset{J}{\tilde f}{}^{mn}\,;\\
&\underset{*N}A^{m}:=\sum\limits_{J=0}^N \epsilon^J\underset{J}A^{m}\,;\quad\underset{*N}j^{m}:=\sum\limits_{J=0}^N \epsilon^J\underset{J}j^{m}\,.
\end{split}
\end{equation}
These are the series representations of the quantities under consideration summed up to a desired order $N$. 

The aim of this section is now to find field equations for the finite series $\underset{*N}{f}{}^{mn}$ respectively $\underset{*N}{A}{}^{n}$. To do so one has to write down some relations for the series of the field invariant $F$ and the pseudo invariant $G$ (compare appendix \ref{ProduktPotenzreihen})
\begin{equation}
\begin{split}\label{PotF}
\left(\underset{*K}F\right)^M=\left(\sum\limits_{Q=0}^K\underset{Q}F\,\epsilon^Q\right)^M&=\sum\limits_{Q=0}^K\underset{Q}{\overset{M}F}\,\epsilon^Q+O\left(\underset{0}{\overset{M}F}\,\epsilon^{M+1}\right)\\
&=\underset{*K}{\left(F^M\right)}+O\left(\underset{0}{F}^M\,\epsilon^{M+1}\right)\,;\\
\quad\text{and}\quad\left(\underset{*K}G\right)^M=\left(\sum\limits_{Q=0}^K\underset{Q}G\,\epsilon^Q\right)^M&=\sum\limits_{Q=0}^K\underset{Q}{\overset{M}G}\,\epsilon^Q+O\left(\underset{0}{\overset{M}F}\,\epsilon^{M+1}\right)\\
&=\underset{*K}{\left(G^M\right)}+O\left(\underset{0}{G}^M\,\epsilon^{M+1}\right)\,.
\end{split}
\end{equation}

At this point we need the series of $L_XY^{mn}$ up to a desired order $M$ in $\epsilon$ entitling it $\underbracket{L_XY^{mn}}_{*M}$. This means we use (\ref{LXY}) but sum up only from $P=0$ to $M$ instead of to infinity. Here we replaced already $\sum\limits_{2N+J=Q}\rightarrow\sum\limits_{N=0}^{Q/2}$ and $J\rightarrow J=Q-2N$ where the summation up to $Q/2$ means that we sum up to the integer part of $Q/2$. This implies that one can replace $Q/2$ by $(Q-1)/2$ in case of odd $Q$
\begin{equation}
\begin{split}
&\underbracket{L_XY^{mn}}_{*M}=\sum\limits_{P=0}^M\sum\limits_{Q=0}^P\sum\limits_{N=0}^{Q/2}\sum\limits_{K=0}^N\sum\limits_{T=0}^{Q-2N}\frac{1}{A^{2N}}\frac{1}{N!}\left(\begin{array}{c}
N\\K
\end{array}\right)\underset{T}{\overset{K}G}\underset{Q-2N-T}{\overset{N-K}F}\times\\
&\quad\qquad\qquad\times\left.\frac{\partial^{N}L_X}{\partial (F/A^2)^{N-K}\partial (G/A^2)^{K}}\right|_{F=0,G=0}\underset{P-Q}Y^{mn}\epsilon^{P-2N}\,.
\end{split}
\end{equation}
Because of $Q\leqq P\leqq M$, $\overset{N}{\underset{J<0}F}\equiv0$ and $\sum\limits_{M=X}^{N<X}\equiv0$ one can rewrite this as
\begin{equation}
\begin{split}
&\underbracket{L_XY^{mn}}_{*M}=\sum\limits_{N=0}^{M/2}\sum\limits_{K=0}^N\frac{1}{N!}\left(\begin{array}{c}
N\\K
\end{array}\right)\left.\frac{\partial^{N}L_X}{\partial (F/A^2)^{N-K}\partial (G/A^2)^{K}}\right|_{F=0,G=0}\times\\
&\quad\qquad\qquad\times\frac{1}{A^{2N}}\sum\limits_{P=0}^M\sum\limits_{Q=0}^P\underset{P-Q}Y^{mn}\epsilon^{P-2N}\sum\limits_{T=0}^{Q-2N}\underset{T}{\overset{K}G}\underset{Q-2N-T}{\overset{N-K}F}\,.
\end{split}
\end{equation}
whereas the sum from $N$ to $Q/2$ was substituted by a sum from $N$ to $M/2$. Comparison of the second line formulas with (\ref{NR2}) ($P\leftrightarrow s$, $2N\leftrightarrow x$, $Q\leftrightarrow t$, $M\leftrightarrow M$ and $Y\leftrightarrow b$, $\sum G F\leftrightarrow a$) yields
\begin{equation}
\begin{split}\label{LXY_A}
&\underbracket{L_XY^{mn}}_{*M}=\sum\limits_{N=0}^{M/2}\sum\limits_{K=0}^N\frac{1}{N!}\left(\begin{array}{c}
N\\K
\end{array}\right)\left.\frac{\partial^{N}L_X}{\partial (F/A^2)^{N-K}\partial (G/A^2)^{K}}\right|_{F=0,G=0}\times\\
&\quad\qquad\qquad\qquad\times\frac{1}{A^{2N}}\left[\left(\sum\limits_{V=0}^{M-2N}\underset{V}Y^{mn}\epsilon^{V}\right)\left(\sum\limits_{V=0}^{M-2N}\epsilon^V\sum\limits_{T=0}^{V}\underset{T}{\overset{K}G}\underset{V-T}{\overset{N-K}F}\right)+\right.\\
&\quad\qquad\qquad\qquad\qquad\qquad\qquad+\left.O\left(\underset{0}{\overset{N}{F}}\underset{0}{Y}^{mn}\epsilon^{M-2N+1}\right)\right]\,.
\end{split}
\end{equation}
Additionally one finds (also compare appendix \ref{ProduktPotenzreihen})
\begin{equation}
\sum\limits_{V=0}^{M-2N}\epsilon^V\sum\limits_{T=0}^{V}\underset{T}{\overset{K}G}\underset{V-T}{\overset{N-K}F}=\left(\sum\limits_{V=0}^{M-2N}\epsilon^V\underset{V}{\overset{K}G}\right)\left(\sum\limits_{V=0}^{M-2N}\epsilon^V\underset{V}{\overset{K}F}\right)+O\left(\underset{0}{\overset{N}{F}}\epsilon^{M-2N+1}\right)\,
\end{equation}
which leads together with (\ref{PotF}) and (\ref{LXY_A}) to 
\begin{equation}
\begin{split}
&\underbracket{L_XY^{mn}}_{*M}=\sum\limits_{N=0}^{M/2}\sum\limits_{K=0}^N\frac{1}{N!}\left(\begin{array}{c}
N\\K
\end{array}\right)\left.\frac{\partial^{N}L_X}{\partial (F/A^2)^{N-K}\partial (G/A^2)^{K}}\right|_{F=0,G=0}\times\\
&\qquad\qquad\qquad\times\frac{1}{A^{2N}}\left[\underset{*(M-2N)}Y^{mn}\left(\underset{*(M-2N)}{F}\right)^{N-K}\left(\underset{*(M-2N)}{G}\right)^{K}\right]\\
&\quad\qquad\qquad\qquad\qquad\qquad\qquad\qquad\qquad\qquad+O\left(\underset{0}{Y}^{mn}\epsilon^{M+1}\right)\,.
\end{split}
\end{equation}
Now it is obvious that for odd (even) $M$ only odd (even) numbered versions of $\underset{*K}Y^{mn}$, $\underset{*K}F$ and $\underset{*K}G$ occur in $\underbracket{L_XY^{mn}}_{*M}$. Additionally one sees immediately that for an even $M$ the mathematical structure of $\underbracket{L_XY^{mn}}_{*M}$ does not change if one replaces $M\rightarrow M+1$ while under this replacement all $\underset{*K}Y^{mn}$, $\underset{*K}F$ and $\underset{*K}G$ are exchanged by $\underset{*K+1}Y^{mn}$, $\underset{*K+1}F$ and $\underset{*K+1}G$. On the other hand the mathematical structure is enriched if one exchanges $M\rightarrow M+2$. This leads to the fact that only the odd numbered versions of $\underbracket{L_XY^{mn}}_{*M}$ are of interest, because the higher order odd (even) numbered solutions are only depending on the lower order odd (even) numbered solutions, but the odd versions secure a higher degree of accurateness at the same stage of mathematical complexity.

At this point it is beneficial to rewrite $\underbracket{L_XY^{mn}}_{*M}$ for odd $M$ and to separate the $\underset{*M}Y^{mn}$ of highest order while we neglect the $O\left(\underset{0}{Y}^{mn}\epsilon^{M+1}\right)$-terms,
\begin{equation}
\begin{split}
\underbracket{L_XY^{mn}}_{*M}\approx\,&\left.L_X\right|_{F=0,G=0}\underset{*M}Y^{mn}+\\
&+\sum\limits_{N=1}^{\frac{M-1}{2}}\sum\limits_{K=0}^N\frac{1}{N!}\left(\begin{array}{c}
N\\K
\end{array}\right)\left.\frac{\partial^{N}L_X}{\partial (F/A^2)^{N-K}\partial (G/A^2)^{K}}\right|_{F=0,G=0}\times\\
&\times\frac{1}{A^{2N}}\left[\underset{*(M-2N)}Y^{mn}\left(\underset{*(M-2N)}{F}\right)^{N-K}\left(\underset{*(M-2N)}{G}\right)^{K}\right]\,.
\end{split}
\end{equation}
Here one has to neglect the second and third lines in case of $M=1$.

Comparison with (\ref{ApproxErregung}) shows that the excitation up to the odd numbered order $M$ is given by
\begin{equation}\label{ErregungenReihe}
\begin{split}
\underset{*M}{h}^{mn}=\,&\sum\limits_{N=0}^M\underset{N}h{}^{mn}\epsilon^N\\
=\,&-2\underbracket{L_FF^{mn}}_{*M}+\underbracket{L_G\tilde{F}{}^{mn}}_{*M}\\
=\,&-2\left.L_F\right|_{F=0,G=0}\underset{*M}f^{mn}+\\
&-2\sum\limits_{N=1}^{\frac{M-1}{2}}\sum\limits_{K=0}^N\frac{1}{N!}\left(\begin{array}{c}
N\\K
\end{array}\right)\left.\frac{\partial^{N}L_F}{\partial (F/A^2)^{N-K}\partial (G/A^2)^{K}}\right|_{F=0,G=0}\times\\
&\times\frac{1}{A^{2N}}\left[\underset{*(M-2N)}f^{mn}\left(\underset{*(M-2N)}{F}\right)^{N-K}\left(\underset{*(M-2N)}{G}\right)^{K}\right]\\
&+\left.L_G\right|_{F=0,G=0}\underset{*M}{\tilde{f}}{}^{mn}+\\
&+\sum\limits_{N=1}^{\frac{M-1}{2}}\sum\limits_{K=0}^N\frac{1}{N!}\left(\begin{array}{c}
N\\K
\end{array}\right)\left.\frac{\partial^{N}L_G}{\partial (F/A^2)^{N-K}\partial (G/A^2)^{K}}\right|_{F=0,G=0}\times\\
&\times\frac{1}{A^{2N}}\left[\underset{*(M-2N)}{\tilde f}{}^{mn}\left(\underset{*(M-2N)}{F}\right)^{N-K}\left(\underset{*(M-2N)}{G}\right)^{K}\right]
\end{split}
\end{equation}

Here a series up to Mth order for the homogeneous (\ref{FG_HOM}) respectively the inhomogeneous (\ref{FG_Inhom}) field equation is needed in analogy to $\underset{*M}{f}^{mn}$, $\underset{*M}{h}^{mn}$ and $\underset{*M}{j}^{m}$. This leads to 
\begin{equation}\label{FeldBisM}
\partial_n\underset{*M}{\tilde{f}}{}^{mn}=0\quad\text{and}\quad\partial_m\underset{*M}{h}{}^{mn}=\frac{4\pi}{c}\,\underset{*M}{j}{}^{n}\,.
\end{equation}
Obviously these field equations have to be fulfilled order by order and has to be solved recursively up the desired order of accuracy. Again the homogeneous equation gives rise to use a potential like in (\ref{Potential_N})
\begin{equation}
\underset{*M}{{f}}{}_{mn}=\partial_m\underset{*M}A{}_n-\partial_n\underset{*M}A{}_m\,.
\end{equation}
From the inhomogeneous field equations for the excitation ($\ref{FeldBisM}_{\mathrm{2}}$) one gets now equations for the field strength series up to Mth order
\begin{equation}\label{GenaeherteFeldgleichungFinal}
\begin{split}
\partial_n\underset{*M}f^{mn}=&\,-\frac{1}{\left.L_F\right|_{F=0,G=0}}\left[\frac{2\pi}{c}\,j^m\right.\\
&\,+\sum\limits_{N=1}^{\frac{M-1}{2}}\sum\limits_{K=0}^N\frac{1}{N!}\left(\begin{array}{c}
N\\K
\end{array}\right)\left.\frac{\partial^{N}L_F}{\partial (F/A^2)^{N-K}\partial (G/A^2)^{K}}\right|_{F=0,G=0}\times\\
&\qquad\times\partial_n\left[\underset{*(M-2N)}f^{mn}\frac{\left(\underset{*(M-2N)}{F}\right)^{N-K}\left(\underset{*(M-2N)}{G}\right)^{K}}{A^{2N}}\right]\\
&-\frac{1}{2}\sum\limits_{N=1}^{\frac{M-1}{2}}\sum\limits_{K=0}^N\frac{1}{N!}\left(\begin{array}{c}
N\\K
\end{array}\right)\left.\frac{\partial^{N}L_G}{\partial (F/A^2)^{N-K}\partial (G/A^2)^{K}}\right|_{F=0,G=0}\times\\
&\qquad\times\partial_n\left.\left[\underset{*(M-2N)}{\tilde f}{}^{mn}\frac{\left(\underset{*(M-2N)}{F}\right)^{N-K}\left(\underset{*(M-2N)}{G}\right)^{K}}{A^{2N}}\right]\right]\,,
\end{split}
\end{equation}
where $\underset{*M}j{}^m$ was exchanged by $j^m$ for the sake of simplicity and in agreement with the procedure of approximation. Analogously it is clear that boundary conditions can be fulfilled if each order fulfills them what eases up the solution of boundary value problems.

Summarizing, we derived a series of linear field equations, one for each field strength series up to $1$st, $3$rd, \ldots, $M$th order, whereat $M$ is an odd positive number. Here the solutions of the lower odd-numbered orders $M-2, M-4,\ldots$ occur as source terms in the field equation for the $M$th order.

However, because the homogeneous field equation is linear one can introduce the Lorentz gauge in all orders simultaneously
\begin{equation}
\partial_m\underset{*M}A{}^m=0\quad\Rightarrow\quad \partial_n\underset{*M}f^{mn}=-\partial_n\partial^n\underset{*M}A{}^m\,.
\end{equation}

To make the situation more lucid we write down the field equations for $M=1$ and $M=3$
\begin{eqnarray}
\label{Ordnung_I}\partial_n\partial^n\underset{*1}A{}^m&=&\frac{1}{\left.L_F\right|_{F=0,G=0}}\frac{2\pi}{c}\,j^m\,;\\
\label{Ordnung_II}\partial_n\partial^n\underset{*3}A{}^m&=&\frac{1}{\left.L_F\right|_{F=0,G=0}}\left[\frac{2\pi}{c}\,j^m\right.\\
\nonumber&&\quad+\partial_n\left(\underset{*1}f^{mn}\left.\frac{\partial L_F}{\partial (F/A^2)}\right|_{F=0,G=0}\frac{\underset{*1}F}{A^{2N}}\right)\\
\nonumber&&\quad+\partial_n\left(\underset{*1}f^{mn}\left.\frac{\partial L_F}{\partial (G/A^2)}\right|_{F=0,G=0}\frac{\underset{*1}G}{A^{2N}}\right)\\
\nonumber&&\quad-\frac12\partial_n\left(\underset{*1}{\tilde f}{}^{mn}\left.\frac{\partial L_G}{\partial (F/A^2)}\right|_{F=0,G=0}\frac{\underset{*1}F}{A^{2N}}\right)\,\\
\nonumber&&\quad\left.-\frac12\partial_n\left(\underset{*1}{\tilde f}{}^{mn}\left.\frac{\partial L_G}{\partial (G/A^2)}\right|_{F=0,G=0}\frac{\underset{*1}G}{A^{2N}}\right)\right]\,,\\
\text{with}\quad\partial_n\underset{*1}A{}^n=0&\,\text{and}\,&\partial_n\underset{*3}A{}^n=0\,.
\end{eqnarray}
Again, the first order recovers Maxwell's original equations in case of $\left.L_F\right|_{F=0,G=0}=-1/2$. From the following orders on one gets equations which have additional source terms build up by the solutions of lower order which describe the self-interaction of the electromagnetic field in this class of theories. This means it is possible to  falsify representatives of the Pleba\'nski class of nonlinear vacuum-electrodynamics by comparing the predictions of its post-Maxwellian terms with experiments. In the next section two examples are considered.

\section{Approximate solution of two static boundary-value problems}\label{section4}
\subsection{Spherical electrostatic source distribution}\label{Kugel}
We consider a homogeneously charged sphere with radius $a$ in spherical coordinates $r,\Theta,\Phi$ where the origin of our coordinates is located in the middle of this sphere. As usual we assume that with $r\rightarrow\infty$ all potentials vanish. 

Looking at (\ref{Ordnung_I}) and the three-dimensional notation in appendix \ref{3DNotation} one sees that in this case the field equations for the potentials up to the first order are given by
\begin{equation}\label{BeispielA}
\begin{split}
&\triangle\underset{*1}\phi=\frac{1}{r^2}\frac{\mathrm d}{\mathrm dr}\left(r^2\,\frac{\mathrm d\underset{*1}\phi}{\mathrm dr}\right)=\frac{1}{\left.L_F\right|_{F=0,G=0}}2\pi\rho(r)=\tilde\rho=\left\{\begin{array}{ll}
\text{const.}&r<a\\
0&r>a\\
\end{array}\right.\,,\\
&\triangle\underset{*1}A^\alpha=0\,,\\
&\text{with}\quad\partial_\alpha\underset{*1}A^\alpha=0\quad\text{and}\quad \partial_t\underset{*1}\phi=0\,.
\end{split}
\end{equation}

Due to the theorem of Kellogg~\cite{Kellogg,Stratton} the magnetic potential and with it the magnetic field strength vanishes in the whole space
\begin{equation}
\underset{*1}A^\alpha=0\quad\Rightarrow\quad\underset{*1}{\mathbf{B}}=0\,.
\end{equation}

Integrating the electrical field equation twice yields ($A,B,C,D$ are constants of integration)
\begin{equation}
\begin{split}
&\frac{\mathrm d\underset{*1}\phi}{\mathrm dr}=\frac{1}{r^2}\int r^2\tilde\rho\,\mathrm dr=\tilde\rho\,\frac{r}3+\frac{A}{r^2}\,\\
\Rightarrow\quad&\underset{*1}\phi=\int\left(\frac{1}{r^2}\int r^2\tilde\rho\,\mathrm dr\right)\mathrm dr=\tilde\rho\,\frac{r^2}{6}-\frac{A}{r}+B\,.
\end{split}
\end{equation}
This leads to the following solution where the total charge of the sphere is given by $Q=4\pi a^3\rho/3$
\begin{equation}
\begin{array}{lll}
\textnormal{Inside:}	& \frac{\mathrm d\underset{*1}\phi}{\mathrm dr}=\frac{Q}{2\left.L_F\right|_{F=0,G=0}a^2}\,\frac{r}{a}+\frac{A}{r^2}\,;	& \underset{*1}\phi=\frac{Q}{2\left.L_F\right|_{F=0,G=0}a^2}\,\frac{r^2}{2a}-\frac{A}{r}+B	\\
\textnormal{Outside:}& \frac{\mathrm d\underset{*1}\phi}{\mathrm dr}=\frac{C}{r^2}\,;	& \underset{*1}\phi=-\frac{C}{r}+D	\\
\end{array}\,.
\end{equation}
Because of $\mathbf E=-\mathrm{grad}\phi=-\mathbf{e}_r\, \mathrm d\phi/\mathrm dr$, the fact that at the origin a singularity is not admitted, the assumption of natural boundary conditions and the necessity to fulfill the jump conditions (compare appendix \ref{Sprung}) on the skin of the sphere one can determine the constants of integration and gets:
\begin{equation}
\begin{array}{llll}
\textnormal{Inside:}&\underset{*1}{\mathbf E}=\mathbf{e}_rE_0\frac{r}{a}\,;	& \frac{\mathrm d\underset{*1}\phi}{\mathrm dr}=-E_0\frac{r}{a}\,;	& \underset{*1}\phi=-E_0\frac{r^2}{2a}+\frac{3}{2}\,E_0a	\\
\textnormal{Outside:}&\underset{*1}{\mathbf E}=\mathbf{e}_rE_0\frac{a^2}{r^2}\,;& \frac{\mathrm d\underset{*1}\phi}{\mathrm dr}=-E_0\frac{a^2}{r^2}\,;	& \underset{*1}\phi=\frac{E_0a^2}{r}\,,	\\
\end{array}
\end{equation}
where $E_0$ is the electrical field strength of first order on the skin of the sphere
\begin{equation}
E_0:=-\frac{Q}{2\left.L_F\right|_{F=0,G=0}a^2}\,.
\end{equation}

Due to (\ref{ErregungenReihe}) and (\ref{Erregung3D}) this results in the following excitations of first order
\begin{equation}
\begin{split}
&\underset{*1}{\mathbf{D}}=-2\left.L_F\right|_{F=0,G=0}\underset{*1}{\mathbf{E}}\,,\\
&\underset{*1}{\mathbf{H}}=-\left.L_G\right|_{F=0,G=0}\underset{*1}{\mathbf{E}}\,.
\end{split}
\end{equation}
Even though the magnetic field strength vanishes everywhere a magnetic excitation exists for theories with $\left.L_G\right|_{F=0,G=0}\neq0$ while parity invariant theories ($L=L(F,G^2)$) always satisfy
\begin{equation}
L_G=2\,\frac{\partial L}{\partial G^2}\,G\quad\Rightarrow\quad\left.L_G\right|_{F=0,G=0}=0\,.
\end{equation}

With knowledge of the solution up to first order the field invariants acting as source terms for the next higher order field equations can be calculated
\begin{equation}
\underset{*1}F=\underset{*1}{\mathbf{B}}^2-\underset{*1}{\mathbf{E}}^2=\left\{\begin{array}{ll}
-E_0^2\frac{r^2}{a^2}&r<a\\
-E_0^2\frac{a^4}{r^4}&r>a\\
\end{array}\right.\quad\text{as well as}\quad \underset{*1}G=0\,.\\
\end{equation}

Now it is possible to calculate the potentials up to the third order whereat the field equations (\ref{Ordnung_II}) read
\begin{equation}\label{BeispielB}
\begin{split}
&\triangle\underset{*3}\phi=-\frac{\left.\frac{\partial L_F}{\partial(F/A^2)}\right|_{F=0,G=0}}{\left.L_F\right|_{F=0,G=0}}\,\mathrm{div}\left(\frac{\underset{*1}{\mathbf{E}}^2}{A^2}\,\underset{*1}{\mathbf{E}}\right)+\frac{1}{\left.L_F\right|_{F=0,G=0}}2\pi\rho(r)\,,\\
&\phantom{\triangle\underset{*3}\phi}=-\frac{\left.\frac{\partial L_F}{\partial(F/A^2)}\right|_{F=0,G=0}}{\left.L_F\right|_{F=0,G=0}}\,E_0^3\left\{\begin{array}{ll}
\phantom-5 (r^2/a^3) & r<a \\ 
-4 (a^6/r^7) & r>a
\end{array} \right.+\frac{1}{\left.L_F\right|_{F=0,G=0}}2\pi\rho(r)\,,\\
&\triangle\underset{*3}A^\alpha=-\frac{\left.\frac{\partial L_G}{\partial(F/A^2)}\right|_{F=0,G=0}}{2\left.L_F\right|_{F=0,G=0}}\,\underset{*1}{\mathbf{E}}\times\mathrm{grad}\frac{\underset{*1}{\mathbf{E}}^2}{A^2}=0\,.
\end{split}
\end{equation}

Comparison of (\ref{BeispielA}) and ($\ref{BeispielB}_{1}$) shows that the solution can be decomposed as $\underset{*3}{\phi}=\underset{*3}{\tilde\phi}+\underset{*1}{\phi}$. Because we have natural boundary conditions both $\underset{*1}{\phi}$ and $\underset{*3}{\tilde\phi}$ are zero for $r\rightarrow\infty$.

Therefore one has only to solve the equations
\begin{equation}\label{Verkuerzt_I}
\begin{split}
&\triangle\underset{*3}{\tilde\phi}=E_0\kappa\left\{\begin{array}{ll}
\phantom-5 (r^2/a^3) & r<a \\ 
-4 (a^6/r^7) & r>a
\end{array} \right.\quad\text{with}\quad\kappa:=-\frac{\left.\frac{\partial L_F}{\partial(F/A^2)}\right|_{F=0,G=0}}{\left.L_F\right|_{F=0,G=0}}\,\frac{E_0^2}{A^2}\,,\\
&\triangle\underset{*3}A^\alpha=0\quad\Rightarrow\quad\underset{*3}{\mathbf{B}}=0\,,
\end{split}
\end{equation}
where the solution is given by ($A,B,C,D$ are constants of integration)
\begin{equation}
\underset{*3}{\tilde\phi}=\left\{\begin{array}{ll}
\frac14\,E_0\kappa\,\frac{r^4}{a^3}-\frac{A}{r}+B & r<a \\ 
-\frac15\,E_0\kappa\,\frac{a^6}{r^5}-\frac{C}{r}+D & r>a
\end{array} \right.\,.
\end{equation}

It remains to determine the constants of integration. Because of $\underset{*3}{\tilde\phi}(r\rightarrow\infty)=0$ and $\underset{*3}{\tilde\phi}(0)\neq\pm\infty$ as well as the requirement of a smooth field strength at the surface of the sphere (compare appendix \ref{Sprung}) one gets
\begin{equation}
A=C=D=0\quad\text{as well as}\quad B=-\frac{9}{20}\,\kappa\,a\,.
\end{equation}
With $\underset{*3}{\phi}=\underset{*3}{\tilde\phi}+\underset{*1}{\phi}$ this leads to
\begin{equation}
\underset{*3}{\mathbf{E}}=-\mathrm{grad}\,\underset{*3}{\phi}=E_0\frac{\mathbf{r}}{r}\left\{\begin{array}{ll}
\frac{r}{a}\left(1-\kappa\,\frac{r^2}{a^2}\right) & r<a \\ 
\frac{a^2}{r^2}\left(1-\kappa\,\frac{a^4}{r^4}\right) & r>a
\end{array} \right.\,.
\end{equation}

In case of theories with $\kappa>0$ -- for example in Born-Infeld's theory -- the correction up to the third order leads to a field strength which is a little bit smaller than the one up to the first order. In the field far away from the source ($r\rightarrow\infty$) this effect vanishes. The new maximum field strength reached on the boundary of the sphere is 
\begin{equation}
\left|E_{\mathrm{max}}\right|=\left|E_0\left(1-\kappa\right)\right|\,.
\end{equation}
It should be underlined here that for all theories of Plebanski's class the calculated solution has a vanishing magnetic field strength in all possible variants but for $\left.L_G\right|_{F=0,G=0}\neq0$  there is a non-vanishing magnetic excitation which possibly can be tested by experiments.

Furthermore we want to underline that $r=a$ is an equipotential surface and one can replace this surface by an ideal conducting foil on the same potential. This means that with the solution of a homogeneously charged sphere also the exterior solution of a spherical charged surface with the charge density of $Q/4\pi\,a^2$ is known. Analogously to Maxwell's theory the field strength inside of this spherical charged surface vanishes because the theorem of Kellogg leads to a constant potential. Here one has to remark that this vanishing field strength up to first order leads to a vanishing source term up to the third order, and therefore also the field strength up to third order vanishes inside.

Hence for electrostatic problems a part of space confined by an equipotential surface is -- like in Maxwell's original theory -- field free in the whole Pleba\'nski class. 

\subsection{Stationary current density along the $z$-axis}\label{Strom}

A constant and homogeneous stationary current density along the $z$-axis in cylindrical coordinates $r,\Phi,z$ is assumed
\begin{equation}
\mathbf{j}=j_z\mathbf{e}_z\quad\text{with}\quad j_z=\left\{\begin{array}{ll}
\text{const.} & r<a \\ 
0 & r>a \\
\end{array} \right.\quad\text{as well as}\quad \rho=0 \,.
\end{equation}
Consequently the field equations for the field strength up to first order (\ref{Ordnung_I}) are\footnote{Here one has to keep in mind that in $\text{curl}\,\text{curl}\mathbf A-\text{grad}\,\text{div}\mathbf A=-\triangle\mathbf A$ the operator $\triangle$ represents the Laplace-operator in Cartesian coordinates only, but it is possible to calculate the Laplace-operator of each Cartesian component $A_i$ in different coordinate systems, e. g. cylindrical coordinates.}
\begin{equation}
\begin{split}
&\triangle \underset{*1}A{}_x=\triangle \underset{*1}A{}_y=\triangle \underset{*1}\phi=0\\
\text{as well as}\quad&\triangle \underset{*1}A{}_z=\frac{1}{r}\frac{\mathrm d}{\mathrm dr}\left(r\,\frac{\mathrm d\underset{*1}A{}_z}{\mathrm dr}\right)=\frac{2\pi\,j_z}{c\left.L_F\right|_{F=0,G=0}}=:\tilde j\,,
\end{split}
\end{equation}
and we assume natural boundary conditions. Then Kellogg's theorem leads to 
\begin{equation}
\underset{*1}A{}_x=\underset{*1}A{}_y=\underset{*1}\phi=0\,
\end{equation}
and the general solution for $\underset{*1}A{}_z$ is ($A,B,C,D$ are constants of integration)
\begin{equation}
\begin{split}
&\text{inside:}\quad\frac{\mathrm d\underset{*1}A{}_z}{\mathrm dr}=\frac12\,\tilde j\,r+\frac{A}{r}\quad\text{as well as}\quad \underset{*1}A{}_z=\frac14\,\tilde j\,r^2+A\,\ln r+B\,,\\
&\text{outside:}\quad\frac{\mathrm d\underset{*1}A{}_z}{\mathrm dr}=\frac{C}{r}\quad\text{as well as}\quad \underset{*1}A{}_z=C\,\ln r+D\,.
\end{split}
\end{equation}
From the boundary conditions, the jump conditions on the current's surface and the assumption that the solution should be nonsingular everywhere one gets for the constants of integration
\begin{equation}
A=D=0\,,\quad B=\frac12\,\tilde j\,a^2\,\ln a-\frac14\,\tilde j\,a^2\,,\quad C=\frac12\,\tilde j\,a^2
\end{equation}
and therefore for the field strength
\begin{equation}
\underset{*1}{\mathbf{E}}=0\quad\text{as well as}\quad\underset{*1}{\mathbf{B}}=\mathrm{curl}\underset{*1}{\mathbf{A}}=-\mathbf{e}_\Phi B_0\left\{\begin{array}{ll}
r/a &\,\, r<a \\ 
a/r &\,\, r>a \\
\end{array} \right.\,.
\end{equation}
Here $B_0$ is defined as the magnetic field strength up to first order on the current density's skin
\begin{equation}
B_0:=-\frac{I}{a\,c\left.L_F\right|_{F=0,G=0}}\quad\text{with}\quad I=\pi\,a^2j_z\,.
\end{equation}

This places us in position to calculate the field invariants up to first order
\begin{equation}
\underset{*1}F=\underset{*1}{\mathbf{B}}^2-\underset{*1}{\mathbf{E}}^2=\left\{\begin{array}{ll}
B_0^2\frac{r^2}{a^2}&r<a\\
B_0^2\frac{a^2}{r^2}&r>a\\
\end{array}\right.\quad\text{and}\quad \underset{*1}G=0\,\\
\end{equation}
and to write down the field equation up to third order (\ref{Ordnung_II})
\begin{equation}
\begin{split}
&\triangle\underset{*3}\phi=-\frac{\left.\frac{\partial L_G}{\partial(F/A^2)}\right|_{F=0,G=0}}{2\left.L_F\right|_{F=0,G=0}}\,\mathrm{div}\left(\frac{\underset{*1}{\mathbf{B}}^2}{A^2}\,\underset{*1}{\mathbf{B}}\right)=0\quad\Rightarrow\quad\underset{*3}\phi=0\,;\quad\underset{*3}{\mathbf{E}}=0\,,\\
&\triangle\underset{*3}{\mathbf A}=\frac{\left.\frac{\partial L_F}{\partial(F/A^2)}\right|_{F=0,G=0}}{\left.L_F\right|_{F=0,G=0}}\,\mathrm{curl}\frac{\underset{*1}{\mathbf{B}}^3}{A^2}+\frac{2\pi\,j_z}{c\left.L_F\right|_{F=0,G=0}}\,.
\end{split}
\end{equation}

Analogously to the foregoing example (\ref{Verkuerzt_I}) we write the solution as the sum $\underset{*3}{{\mathbf A}}=\underset{*3}{\tilde{\mathbf A}}+\underset{*1}{{\mathbf A}}$ and calculate only $\underset{*3}{\tilde{\mathbf A}}$. To do so we rearrange at first the right side of the field equation 
\begin{equation}
\mathrm{curl}\,\frac{\underset{*1}{\mathbf{B}}^3}{A^2}=\frac{\mathbf{e}_z}{r}\frac{\mathrm d}{\mathrm dr}\frac{r\,\underset{*1}{B}{}_\Phi^3}{A^2}=\mathbf{e}_z\frac{B_0^3}{A^2}\left\{\begin{array}{ll}
4\frac{r^2}{a^3}&r<a\\
-2\frac{a^3}{r^4}&r>a\\
\end{array}\right.\,.
\end{equation}
and get therefore:
\begin{equation}
\begin{split}
&\triangle\underset{*3}{\tilde A_x}=\triangle\underset{*3}{\tilde A_y}=0\quad\Rightarrow\quad\underset{*3}{\tilde A_x}=\underset{*3}{\tilde A_y}=0\,,\\
&\triangle\underset{*3}{\tilde A_z}=\frac{1}{r}\frac{\mathrm d}{\mathrm dr}\left(r\,\frac{\mathrm d\underset{*3}{\tilde A_z}}{\mathrm dr}\right)=\frac{\left.\frac{\partial L_F}{\partial(F/A^2)}\right|_{F=0,G=0}}{\left.L_F\right|_{F=0,G=0}}\,\frac{B_0^3}{A^2}\left\{\begin{array}{ll}
4\frac{r^2}{a^3}&r<a\\
-2\frac{a^3}{r^4}&r>a\\
\end{array}\right.\,.
\end{split}
\end{equation}
The solution is 
\begin{equation}
\begin{split}
&\frac{\underset{*3}{\mathrm d\tilde A_z}}{\mathrm dr}=\left\{\begin{array}{ll}
B_0\sigma\,\frac{r^3}{a^3}+\frac{A}{r}&r<a\\
B_0\sigma\,\frac{a^3}{r^3}+\frac{C}{r}&r>a\\
\end{array}\right.\,,\\
&\underset{*3}{\tilde A_z}=\left\{\begin{array}{ll}
\frac14\,B_0\sigma\,\frac{r^4}{a^3}+A\ln r+B&r<a\\
-\frac12\,B_0\sigma\,\frac{a^3}{r^2}+C\ln r+D&r>a\\
\end{array}\right.\,,
\end{split}
\end{equation}
where the constant $\sigma$ is introduced
\begin{equation}
\sigma:=\frac{\left.\frac{\partial L_F}{\partial(F/A^2)}\right|_{F=0,G=0}}{\left.L_F\right|_{F=0,G=0}}\,\frac{B_0^2}{A^2}\,.
\end{equation}
Again the boundary conditions and the jump conditions on the current's surface (compare appendix \ref{Sprung}) as well as the assumption that the solution should be nonsingular everywhere determine the constants of integration
\begin{equation}
A=C=D=0\,;\quad B=-\frac34\,B_0\sigma\,a\,.
\end{equation}
Thus one obtains for the field strength up to third order
\begin{equation}\label{85}
\underset{*3}{\mathbf{B}}=\mathrm{curl}\left(\underset{*1}{\mathbf{A}}+\underset{*3}{\tilde{\mathbf{A}}}\right)=\underset{*1}{\mathbf{B}}-\mathbf{e}_z\frac{\mathrm d\underset{*3}{\tilde A_z}}{\mathbf dr}=-\mathbf{e}_\Phi B_0\left\{\begin{array}{ll}
r/a\left(1+\sigma\,\frac{r^2}{a^2}\right) & r<a \\ 
a/r\left(1+\sigma\,\frac{a^2}{r^2}\right) & r>a \\
\end{array} \right.\,.
\end{equation}
This means that for all theories with $\sigma<0$, like Born-Infeld's theory, the field strength is smaller than in Maxwell's theory. Accordingly to (\ref{85}) the maximum value which is reached on the skin of the current density is generally given by
\begin{equation}
\left|B_{\mathrm{max}}\right|=\left|B_0\left(1+\sigma\right)\right|\,.
\end{equation}

Analogously to our electrostatic example the field-generating configuration does not lead to any electrostatic field strength for the whole Pleba\'nski class. However theories with $\left.L_G\right|_{F=0,G=0}\neq0$ lead to an electrostatic excitation because of (\ref{Erregung_Test}) (compare also (\ref{Erregung3D})).

Completing this we want to state an analogue to the foregoing electrostatic example. Here the magnetic field is derived from an item ($A_z$) behaving in our context like a kind of scalar potential 
\begin{equation}
\mathbf{B}=\mathrm{curl}\,\mathbf{A}=-\mathbf{e}_z\times\mathrm{grad}\,A_z\,.
\end{equation}
Because the skin of the $z$-directed conductor is an equipotential surface of $A_z$ one can substitute this surface by a perfect conducting foil with the constant surface current $I/2\pi\,a$. Hence the calculated solution is also valid for the exterior domain of a coaxial conductor of radius $a$ flowed through by the current $I$. Analogue to the electrostatic case the magnetic field inside vanishes as a consequence of Kellogg's theorem.  

\section{Conclusion}\label{section5}
 
With the aid of a weak field strength approximation (\ref{analog}) in $\epsilon\sim F^{mn}/A$ we calculated a series expansion for the excitation (\ref{ApproxErregung}) for the whole Pleba\'nski class of nonlinear electrodynamics. Furthermore we developed a set of linear field equations one for each order of approximation, where these field equations can be solved order by order up to the desired degree of accuracy (sections \ref{TheHomogeneousMaxwellEquation} and \ref{TheInhomogeneousMaxwellEquation}). Here the solutions of the lower orders act as additional source terms of the higher orders and represent the "post-Maxwellian" self-interaction of the nonlinear electrodynamical field.

At this point it is remarkable that each order of even number enriches the mathematical structure of the approximate field equation's source terms by adding higher powers of the field invariants built up by the solutions of lower order. On the other hand each following odd numbered order does not enrich this structure but rather completes it to some extent. 

In section \ref{CombinedFormOfApproximateFieldEquations} it was shown that one can reformulate the approximate field equations in a way that not only one order of the field strength but the whole series up to a desired order can be calculated by one of the field equations (\ref{GenaeherteFeldgleichungFinal}). Again, the solutions of the lower orders ($<N$) lead to additional source terms in the field equation of order $N$ as well as the solution of order $N$ comes into play as source term in the orders above ($>N$). Here we found that it is sufficient to calculate only the solutions up to the odd numbered orders what makes the field equation for the solution of the even numbered orders unnecessary. 

Only in the special case of $\left.L_F\right|_{F=0,G=0}=-1/2$ Maxwell's original theory can be recovered in the lowest order. Here one has to decide between situations with and without external sources $j^m$. In case of $j^m=0$ Maxwell's original theory occurs not only for low but also for arbitrary frequencies as limiting case for weak fields (compare (\ref{tau2})).

Our approach makes it also possible to calculate "post-Maxwellian" corrections for problems where the solution of Maxwell's original equations is bounded, i.e. nonsingular. If this is not fulfilled everywhere one can restrict to some area where it is. One sees for example that the far field of a Hertzian dipole in Maxwell's theory has vanishing field invariants so that the far field in Pleba\'nski's class has only vanishing "post-Maxwellian" terms. 

The calculations of "post-Maxwellian" corrections offer new possibilities for experimental tests of Pleba\'nski's class. Moreover, if it should be possible to measure the excitation of the nonlinear field directly this should enable one to decide between parity and  non parity invariant theories, because $\left.L_F\right|_{F=0,G=0}=-1/2$ leeds to Maxwell's original field equations for the field strength in the first order\footnote{Compare (\ref{Erregung_Test}) and below as well as (\ref{Ordnung_II}).}, but does not recover Maxwell's consitutive law for the vacuum ($H^{mn}=F^{mn}$) necessarily. Instead the first order constitutive law reads $\underset{*1}h^{mn}=\underset{*1}{f}{}^{mn}+\left.L_G\right|_{F=0,G=0}\underset{*1}{\tilde f}{}^{mn}$ which could give rise for further experimental tests.

To give two examples the first "post-Maxwellian" corrections for a homogeneously charged sphere (\ref{Kugel}) and for a homogeneous linear thick stream current (\ref{Strom}) were calculated. Here one sees that the field is diminished or amplified near by the source while far away this effect vanishes where a criterion is formulated making decidable which case comes to pass for a given theory. As one would expect in Born-Infeld's case the maximum field strength is lowered. 

Finally it should be remarked that it is possible to use the approximation procedure developed here also in terms of the excitation instead of the field strength if one confines oneself to the vacuum theory. To go over to the corresponding field equations one can use the scheme presented in~\cite{Michelson}. If this scheme is used also for the approximate field equations one gets the whole approximation method in terms of the excitation, but one should keep in mind that the potential of the field strength becomes an anti-potential of the excitation.

\section*{Acknowledgments}
I gratefully acknowledge a PhD stipend from Evanglisches Studienwerk Villigst during this work as well as support from the Deutsche Forschungsgemeinschaft within the Research Training Group 1620 "Models of Gravity." Furthermore I have to thank H. v. Borzeszkowski most sincerely for helpful discussions and comments. 

\appendix
\section*{Appendix}
\addcontentsline{toc}{section}{Appendices}
\renewcommand{\thesubsection}{\Alph{subsection}}
\numberwithin{equation}{subsection}
\subsection{Products of power series}\label{ProduktPotenzreihen}
Two power series -- in our context also known as "Einstein series" for example used for the Einstein-Infeld-Hoffmann approximation\footnote{For a lucid representation compare~\cite{Bergmann}.} -- can be multiplied as follows
\begin{equation}
\begin{split}
&f(\epsilon):=\sum\limits_{s=0}^\infty a_s\epsilon^s\,;\quad g(\epsilon):=\sum\limits_{s=0}^\infty b_s\epsilon^s\\
\Rightarrow\quad&f(\epsilon)\cdot g(\epsilon)=\sum\limits_{s=0}^\infty\left(\sum\limits_{t=0}^sa_tb_{s-t}\right)\epsilon^s\,.
\end{split}
\end{equation}

If these series are restricted to a desired order $M$ one gets instead
\begin{equation}
\begin{split}
&f(\epsilon,M):=\sum\limits_{s=0}^M a_s\epsilon^s\,;\quad g(\epsilon,M):=\sum\limits_{s=0}^M b_s\epsilon^s\\
\Rightarrow\quad&f(\epsilon,M)\cdot g(\epsilon,M)=\sum\limits_{s=0}^M\left(\sum\limits_{t=0}^sa_tb_{s-t}\right)\epsilon^s+O(a_0b_0\epsilon^{M+1})\,.
\end{split}
\end{equation}
Here $O(a_0b_0\epsilon^{M+1})$ represents terms of higher order. Additionally one has to keep in mind that the sorting by a parameter of smallness ($\epsilon$) makes sense only if $a_s\sim  b_s\sim a_0\sim b_0$ holds.

A further type of series is needed in our context ($a_{<0}\equiv0$; $t,s,x$ are integers; $v:=t-x$; $Q:=s-x$)
\begin{eqnarray}\label{NR2}
\nonumber\sum\limits_{s=0}^M\sum\limits_{t=0}^sa_{t-x}b_{s-t}\,\epsilon^{s-x}&=&\sum\limits_{s=x}^M\sum\limits_{t=x}^sa_{t-x}b_{s-t}\,\epsilon^{s-x}=\sum\limits_{s=x}^M\sum\limits_{v=0}^{s-x}a_{v}b_{s-v-x}\,\epsilon^{s-x}\\
&=&\sum\limits_{Q=0}^{M-x}\sum\limits_{v=0}^Qa_{v}b_{Q-v}\,\epsilon^{Q}\\
\nonumber&=&\left(\sum\limits_{Q=0}^{M-x}a_{Q}\,\epsilon^{Q}\right)\left(\sum\limits_{Q=0}^{M-x}b_{Q}\,\epsilon^{Q}\right)+O(a_0b_0\epsilon^{M-x+1})\,.
\end{eqnarray}

\subsection{Powers of power series}\label{PotenzenVonPotenzreihen}
Here we follow the approach described in~\cite{Potenzreihen} adapted to our situation. We assume that a power series in $\epsilon$ is given
\begin{equation}
g(\epsilon)=\sum\limits_{J=0}^\infty g_J\epsilon^J\,.
\end{equation}
and wish to calculate their powers in the domain of natural numbers also as a power series in $\epsilon$
\begin{equation}
G(\epsilon):=\left(g(\epsilon)\right)^N=\left(\sum\limits_{J=0}^\infty g_J\epsilon^J\right)^N=\sum\limits_{J=0}^\infty G_J\epsilon^J\,.
\end{equation}
To do so one has to calculate the coefficient $G_J$ as follows
\begin{equation}
G_J=\frac{1}{J!}\left.\frac{\mathrm{d}^J}{\mathrm{d}g^J}\,g^N\right|_{\epsilon=0}\,.
\end{equation}
To calculate this derivative one applies Fa{\'a} di Bruno's formula which makes it possible to calculate higher derivatives of composed functions such as $F(t)=F(f(t))$:
\begin{equation}
\frac{\mathrm{d}^J F(f(t))}{\mathrm{d}t^J}=\sum\limits_{s=1}^S\frac{J!}{k_{1_s}!\cdots\ldots\cdot k_{J_s}!}\,\frac{\mathrm{d}^M F}{\mathrm{d}f^M}\,\prod\limits_{L=1}^{J}\left(\frac{1}{L!}\frac{\mathrm{d}^Lf(t)}{\mathrm{d}t^L}\right)^{k_{L_s}}\,.
\end{equation}
Here the summation runs over all solutions of the following formula where $s$ counts the solutions and $S$ is the total number of solutions
\begin{equation}
k_{1s}+2k_{2s}+\ldots+Jk_{Js}=J\,.
\end{equation}
Additionally $M$ is defined as follows
\begin{equation}
k_{1_s}+k_{2_s}+\ldots+k_{J_s}=:M\,.
\end{equation}

This can now be used to calculate $G_J$ as
\begin{equation}\label{KoeffPotPot}
G_J=\sum\limits_{s=1}^S\frac{N(N-1)\cdot\ldots\cdot(N-M+1)}{k_{1_s}!\cdots\ldots\cdot k_{J_s}!}\,g_0^{N-M}g_1^{k_{1_s}}\cdot\ldots\cdot g_J^{k_{J_s}}\,.
\end{equation}
with again
\begin{equation}
k_{1s}+2k_{2s}+\ldots+Jk_{Js}=J\quad\text{and}\quad k_{1_s}+k_{2_s}+\ldots+k_{J_s}=M\leqq N\,,
\end{equation}
where $M\leqq N$ has to be fulfilled because otherwise the corresponding $G_J$ vanishes identically.

\subsection{Three-dimensional notation}\label{3DNotation}
For the discussion of the two examples in section \ref{section4} a three-dimensional decomposition of the field equations, etc. is needed. Therefore we give here the relations between the three- and four-dimensional quantities in Cartesian coordinates where as usual $E_\alpha$ denotes the electric field strength, $D_\alpha$ the electric excitation, $B_\alpha$ the magnetic field strength and $H_\alpha$ the magnetic excitation ($\varepsilon^{123}=1$)
\begin{equation}\label{eq:EB}
\begin{split}
&E_{\alpha}=F_{\alpha 4}\,,\quad B^\alpha=\frac12\,\varepsilon^{\alpha\beta\gamma}F_{\beta\gamma}=-\tilde{F}{}^{\alpha 4}\,,\\
&D^{\alpha}=-H^{\alpha 4}\,,\quad H_{\alpha}=\frac12\,\varepsilon_{\alpha\beta\gamma}H^{\beta\gamma}=\tilde{H}{}_{\alpha 4}\,.
\end{split}
\end{equation}
For the invariants this leads to
\begin{eqnarray}
F=\frac{1}{2}\,F_{mn}F^{mn}=\mathbf B^2-\mathbf E^2\quad&\text{and}&\quad G=-\frac{1}{4}\,F_{mn}\tilde F^{mn}=\mathbf B\mathbf E\,.
\end{eqnarray}
Therefore the constitutive law for the vacuum can be rewritten in three-dimensional notation
\begin{equation}
\begin{split}
&D^\iota=H^{4\iota}=-\frac{\partial  L}{\partial F_{4\iota}}=\frac{\partial  L}{\partial E_\iota}\,,\\
&H_\iota=\frac12\,\varepsilon_{\iota\alpha\beta} H^{\alpha\beta}=-\frac12\,\varepsilon_{\iota\alpha\beta}\frac{\partial  L}{\partial F_{\alpha\beta}}=-\frac{\partial  L}{\partial B^\iota}\,,
\end{split}
\end{equation}
where for the Pleba\'nski class with the Lagrangian $L=L(F,G)$ this simplifies to
\begin{equation}\label{Erregung3D}
\begin{split}
&D_\iota=-2E_\iota\,L_F+B_\iota\,L_G\,,\\
&H_\iota=-2B_\iota\,L_F-E_\iota\,L_G\,.
\end{split}
\end{equation}
Further we decompose the current density $j^m$ into
\begin{equation}
(j^m)=(j^\alpha,c\rho)\,
\end{equation}
and the potential into 
\begin{equation}
(A_a)=\left(A_\alpha,-\phi\right)\,.
\end{equation}

This leads to Maxwell's equations in three-dimensional notation
\begin{equation}\label{Maxwell_Glg_3D}
\begin{split}
&\text{curl}\,\mathbf H=\frac{1}{c}\frac{\partial \mathbf D}{\partial t}+\frac{4\pi}{c}\mathbf j\,;\quad \text{div}\,\mathbf D=4\pi\rho\,;\\
&\text{curl}\,\mathbf E=-\frac{1}{c}\frac{\partial \mathbf B}{\partial t}\,;\quad\text{div}\,\mathbf B=0\,.
\end{split}
\end{equation}
Again, the homogeneous equations of the second line above are identically fulfilled due to the use of the decomposed potential 
\begin{equation}
\begin{split}
F_{mn}=\partial_mA_n-\partial_nA_m\quad\Leftrightarrow\quad\mathbf E=-\text{grad}\,\phi-\frac{1}{c}\,\frac{\partial\mathbf A}{\partial t}\,;\quad\mathbf B=\text{curl}\,\mathbf A\,.
\end{split}
\end{equation}

\subsection{Jump conditions of the electromagnetic field at interfaces}\label{Sprung}
From the usual formulation of Maxwell's equations for field strength and excitation (\ref{Maxwell_Glg_3D}) the jump conditions at interfaces are given by (compare for example~\cite{Stratton})
\begin{equation}\label{Sprung2}
\begin{split}
\mathbf n\cdot(\mathbf B_2-\mathbf B_1)=0\,;&\qquad\mathbf n\cdot(\mathbf D_2-\mathbf D_1)=\lambda\,;\\
\mathbf n\times(\mathbf E_2-\mathbf E_1)=0\,;&\qquad\mathbf n\times(\mathbf H_2-\mathbf H_1)=\mathbf K/c\,,
\end{split}
\end{equation}
where $\mathbf{n}$ denotes the surface normal looking into the second half space. Additionally $\lambda$ is the surface density of charge where $\mathbf K$ is the surface current on the interface. 

This can also be formulated in covariant notation where the equation of the interface (a three dimensional hypersurface) is given by
\begin{equation}
Z(x^m)=0\,.
\end{equation}
Building the four normal of $Z$ yields ($\overset{R}=$ denotes the rest system) 
\begin{equation}
N_n=\frac{\partial_n Z}{\sqrt{g^{ab}\partial_a Z\partial_b Z}}\overset{R}=\left(\mathbf n\,,\,\,0\right)\quad\text{with}\quad\partial_nZ(x^m)=\left(\mathrm{grad}Z,\frac{\partial Z}{c\partial t}\right)\,.
\end{equation}

The conditions (\ref{Sprung2}) can then be interpreted as the rest system representation of the following covariant conditions~\cite{Stephani}
\begin{eqnarray}
N_n\left(\tilde F_{\text{2}}^{mn}-\tilde F_{\text{1}}^{mn}\right)=0\,;&&N_n\left(H_{\text{2}}^{mn}-H_{\text{1}}^{mn}\right)=\frac{1}{c}\,j^m_{\text{surface}}\,,
\end{eqnarray}
where the vector of the four surface current becomes in the rest system 
\begin{equation}
j^m_{\text{surface}}\overset{R}=\left(\mathbf K\,,\,\,c\lambda\right)\,.
\end{equation}

\subsection{Fourth and fifth oder of the excitation series}\label{FuenfteVierteOrd}
\allowdisplaybreaks To keep the arrangement clear we write down the fourth and the fifth orders of the excitation series (\ref{ApproxErregung}) here in the appendix
\begin{align}
\nonumber\epsilon^4\underset{4}h^{mn}=\,&\epsilon^4\left(-2\left.L_F\right|_{F=0,G=0}\underset{4}{f}{}^{mn}+\left.L_G\right|_{F=0,G=0}\underset{4}{\tilde f}{}^{mn}\right)\\
\nonumber&-2\epsilon^2\left(\frac{\underset{0}F}{A^2}\left.\frac{\partial L_F}{\partial (F/A^2)}\right|_{F=0,G=0}+\frac{\underset{0}G}{A^2}\left.\frac{\partial L_F}{\partial (G/A^2)}\right|_{F=0,G=0}\right)\underset{2}{f}{}^{mn}\\
\nonumber&+\epsilon^2\left(\frac{\underset{0}F}{A^2}\left.\frac{\partial L_G}{\partial (F/A^2)}\right|_{F=0,G=0}+\frac{\underset{0}G}{A^2}\left.\frac{\partial L_G}{\partial (G/A^2)}\right|_{F=0,G=0}\right)\underset{2}{\tilde f}{}^{mn}\\
\nonumber&-2\epsilon^2\left(\frac{\underset{1}F}{A^2}\left.\frac{\partial L_F}{\partial (F/A^2)}\right|_{F=0,G=0}+\frac{\underset{1}G}{A^2}\left.\frac{\partial L_F}{\partial (G/A^2)}\right|_{F=0,G=0}\right)\underset{1}{f}{}^{mn}\\
\nonumber&+\epsilon^2\left(\frac{\underset{1}F}{A^2}\left.\frac{\partial L_G}{\partial (F/A^2)}\right|_{F=0,G=0}+\frac{\underset{1}G}{A^2}\left.\frac{\partial L_G}{\partial (G/A^2)}\right|_{F=0,G=0}\right)\underset{1}{\tilde f}{}^{mn}\\
\nonumber&-2\epsilon^2\left(\frac{\underset{2}F}{A^2}\left.\frac{\partial L_F}{\partial (F/A^2)}\right|_{F=0,G=0}+\frac{\underset{2}G}{A^2}\left.\frac{\partial L_F}{\partial (G/A^2)}\right|_{F=0,G=0}\right)\underset{0}{f}{}^{mn}\\
\nonumber&+\epsilon^2\left(\frac{\underset{2}F}{A^2}\left.\frac{\partial L_G}{\partial (F/A^2)}\right|_{F=0,G=0}+\frac{\underset{2}G}{A^2}\left.\frac{\partial L_G}{\partial (G/A^2)}\right|_{F=0,G=0}\right)\underset{0}{\tilde f}{}^{mn}\\
\nonumber&-2\,\frac12\left(\frac{\underset{0}{\overset{2}F}}{A^4}\left.\frac{\partial^2 L_F}{\partial (F/A^2)^2}\right|_{F=0,G=0}+2\frac{\underset{0}{\overset{1}F}\underset{0}{\overset{1}G}}{A^4}\left.\frac{\partial^2 L_F}{\partial (F/A^2)\partial (G/A^2)}\right|_{F=0,G=0}\right.\\
\nonumber&\qquad\qquad\qquad\qquad\qquad\qquad+\left.\frac{\underset{0}{\overset{2}G}}{A^4}\left.\frac{\partial^2 L_F}{\partial (G/A^2)^2}\right|_{F=0,G=0}\right)\underset{0}{f}{}^{mn}\\
\nonumber&+\frac12\left(\frac{\underset{0}{\overset{2}F}}{A^4}\left.\frac{\partial^2 L_G}{\partial (F/A^2)^2}\right|_{F=0,G=0}+2\frac{\underset{0}{\overset{1}F}\underset{0}{\overset{1}G}}{A^4}\left.\frac{\partial^2 L_G}{\partial (F/A^2)\partial (G/A^2)}\right|_{F=0,G=0}\right.\\
\nonumber&\qquad\qquad\qquad\qquad\qquad\qquad+\left.\frac{\underset{0}{\overset{2}G}}{A^4}\left.\frac{\partial^2 L_G}{\partial (G/A^2)^2}\right|_{F=0,G=0}\right)\underset{0}{\tilde f}{}^{mn}\,
\end{align}
and
\begin{align}
\nonumber\epsilon^5\underset{5}h^{mn}=\,&\epsilon^5\left(-2\left.L_F\right|_{F=0,G=0}\underset{5}{f}{}^{mn}+\left.L_G\right|_{F=0,G=0}\underset{5}{\tilde f}{}^{mn}\right)\\
\nonumber&-2\epsilon^3\left(\frac{\underset{0}F}{A^2}\left.\frac{\partial L_F}{\partial (F/A^2)}\right|_{F=0,G=0}+\frac{\underset{0}G}{A^2}\left.\frac{\partial L_F}{\partial (G/A^2)}\right|_{F=0,G=0}\right)\underset{3}{f}{}^{mn}\\
\nonumber&+\epsilon^3\left(\frac{\underset{0}F}{A^2}\left.\frac{\partial L_G}{\partial (F/A^2)}\right|_{F=0,G=0}+\frac{\underset{0}G}{A^2}\left.\frac{\partial L_G}{\partial (G/A^2)}\right|_{F=0,G=0}\right)\underset{3}{\tilde f}{}^{mn}\\
\nonumber&-2\epsilon^3\left(\frac{\underset{1}F}{A^2}\left.\frac{\partial L_F}{\partial (F/A^2)}\right|_{F=0,G=0}+\frac{\underset{1}G}{A^2}\left.\frac{\partial L_F}{\partial (G/A^2)}\right|_{F=0,G=0}\right)\underset{2}{f}{}^{mn}\\
\nonumber&+\epsilon^3\left(\frac{\underset{1}F}{A^2}\left.\frac{\partial L_G}{\partial (F/A^2)}\right|_{F=0,G=0}+\frac{\underset{1}G}{A^2}\left.\frac{\partial L_G}{\partial (G/A^2)}\right|_{F=0,G=0}\right)\underset{2}{\tilde f}{}^{mn}\\
\nonumber&-2\epsilon^3\left(\frac{\underset{2}F}{A^2}\left.\frac{\partial L_F}{\partial (F/A^2)}\right|_{F=0,G=0}+\frac{\underset{2}G}{A^2}\left.\frac{\partial L_F}{\partial (G/A^2)}\right|_{F=0,G=0}\right)\underset{1}{f}{}^{mn}\\
\nonumber&+\epsilon^3\left(\frac{\underset{2}F}{A^2}\left.\frac{\partial L_G}{\partial (F/A^2)}\right|_{F=0,G=0}+\frac{\underset{2}G}{A^2}\left.\frac{\partial L_G}{\partial (G/A^2)}\right|_{F=0,G=0}\right)\underset{1}{\tilde f}{}^{mn}\\
\nonumber&-2\epsilon^3\left(\frac{\underset{3}F}{A^2}\left.\frac{\partial L_F}{\partial (F/A^2)}\right|_{F=0,G=0}+\frac{\underset{3}G}{A^2}\left.\frac{\partial L_F}{\partial (G/A^2)}\right|_{F=0,G=0}\right)\underset{0}{f}{}^{mn}\\
\nonumber&+\epsilon^3\left(\frac{\underset{3}F}{A^2}\left.\frac{\partial L_G}{\partial (F/A^2)}\right|_{F=0,G=0}+\frac{\underset{3}G}{A^2}\left.\frac{\partial L_G}{\partial (G/A^2)}\right|_{F=0,G=0}\right)\underset{0}{\tilde f}{}^{mn}\\
\nonumber&-2\,\frac12\epsilon\left(\frac{\underset{0}{\overset{2}F}}{A^4}\left.\frac{\partial^2 L_F}{\partial (F/A^2)^2}\right|_{F=0,G=0}+2\frac{\underset{0}{\overset{1}F}\underset{0}{\overset{1}G}}{A^4}\left.\frac{\partial^2 L_F}{\partial (F/A^2)\partial (G/A^2)}\right|_{F=0,G=0}\right.\\
\nonumber&\qquad\qquad\qquad\qquad\qquad\qquad+\left.\frac{\underset{0}{\overset{2}G}}{A^4}\left.\frac{\partial^2 L_F}{\partial (G/A^2)^2}\right|_{F=0,G=0}\right)\underset{1}{f}{}^{mn}\\
\nonumber&+\frac12\epsilon\left(\frac{\underset{0}{\overset{2}F}}{A^4}\left.\frac{\partial^2 L_G}{\partial (F/A^2)^2}\right|_{F=0,G=0}+2\frac{\underset{0}{\overset{1}F}\underset{0}{\overset{1}G}}{A^4}\left.\frac{\partial^2 L_G}{\partial (F/A^2)\partial (G/A^2)}\right|_{F=0,G=0}\right.\\
\nonumber&\qquad\qquad\qquad\qquad\qquad\qquad+\left.\frac{\underset{0}{\overset{2}G}}{A^4}\left.\frac{\partial^2 L_G}{\partial (G/A^2)^2}\right|_{F=0,G=0}\right)\underset{1}{\tilde f}{}^{mn}\\
\nonumber&-2\epsilon\,\frac12\left(\frac{\underset{1}{\overset{2}F}}{A^4}\left.\frac{\partial^2 L_F}{\partial (F/A^2)^2}\right|_{F=0,G=0}+2\frac{\underset{1}{\overset{1}F}\underset{0}{\overset{1}G}}{A^4}\left.\frac{\partial^2 L_F}{\partial (F/A^2)\partial (G/A^2)}\right|_{F=0,G=0}\right.\\
\nonumber&\;+2\frac{\underset{0}{\overset{1}F}\underset{1}{\overset{1}G}}{A^4}\left.\frac{\partial^2 L_F}{\partial (F/A^2)\partial (G/A^2)}\right|_{F=0,G=0}+\left.\frac{\underset{1}{\overset{2}G}}{A^4}\left.\frac{\partial^2 L_F}{\partial (G/A^2)^2}\right|_{F=0,G=0}\right)\underset{0}{f}{}^{mn}\\
\nonumber&+\epsilon\,\frac12\left(\frac{\underset{1}{\overset{2}F}}{A^4}\left.\frac{\partial^2 L_G}{\partial (F/A^2)^2}\right|_{F=0,G=0}+2\frac{\underset{1}{\overset{1}F}\underset{0}{\overset{1}G}}{A^4}\left.\frac{\partial^2 L_G}{\partial (F/A^2)\partial (G/A^2)}\right|_{F=0,G=0}\right.\\
\nonumber&\;+2\frac{\underset{0}{\overset{1}F}\underset{1}{\overset{1}G}}{A^4}\left.\frac{\partial^2 L_G}{\partial (F/A^2)\partial (G/A^2)}\right|_{F=0,G=0}+\left.\frac{\underset{1}{\overset{2}G}}{A^4}\left.\frac{\partial^2 L_G}{\partial (G/A^2)^2}\right|_{F=0,G=0}\right)\underset{0}{\tilde f}{}^{mn}\,.
\end{align}

\end{document}